\documentclass[aps,prb,reprint,superscriptaddress,amsmath,amssymb]{revtex4-1}
\usepackage{epsfig,psfrag}
\usepackage{dcolumn}
\usepackage{bm}
\usepackage{graphicx}
\usepackage{color}

\begin{document}
\title{Free carrier absorption in cascade structures due to static scatterers in the in-plane polarization}

\author{C. Ndebeka-Bandou}
\author{F.Carosella}
\author{R. Ferreira}
\author{G. Bastard}
\affiliation{Laboratoire Pierre Aigrain, Ecole Normale Sup\'erieure, CNRS (UMR 8551), Universit\'e P. et M. Curie,
Universit\'e Paris Diderot, 24 rue Lhomond F-75005 Paris, France}

\begin{abstract}
We report on the computation of the free carrier absorption induced by static scatterers in cascade structures when the electromagnetic wave propagates along the growth axis. We find that a Drude-like tail exists for this polarization. The absorption is found larger than when the wave propagates in the layer plane. Also intra-subband scattering is found more efficient than inter-subband scattering. The alloy scattering is found to be particularly efficient.  
\end{abstract}


\maketitle

\section{Introduction}

One of the possible ways to realize THz sources is the development of Quantum Cascade Lasers (QCLs) \cite{faist1994,gmachl2001,kohler2002}. However, so far, no room temperature operation has ever been reported and the search for improvement is intensive \cite{walther2007,williams2007,kumar2009}. The Free Carrier Absorption (FCA) is a plausible source of losses for far-IR and THz lasers \cite{vurgaftman1999,faist2007,wacker2011}. It consists in the reabsoprtion of the laser photons by the free carriers \cite{dumke1961}, in particular those that occupy the upper laser subband. FCA arises from intra-subband and inter-subband oblique transitions\cite{ndebeka2012} (in the $\vec{k}$ space) activated by static scatterers or phonons. Recently, we proposed a modeling of FCA in QCL's \cite{carosella2012}, thereby focusing our attention on the standard light polarization for these systems, i.e, electromagnetic waves that propagate in the layer plane with their electric vector parallel to the growth axis. In this paper we examine the case of an electromagnetic wave that propagates along the ($z$) growth axis with its electric vector laying in the layer plane. We shall show that FCA exibits a Drude-like behavior for this electromagnetic polarization in stark contrast with the previous results. We shall also demonstrate that FCA can be larger in the in-plane polarization that in the $z$-polarization. Finally, we shall show that the intra-subband FCA is more important than the inter-suband one, a feature that again differs from what happens in the $z$-polarization analyzed previously\cite{carosella2012}.

In the following, we concentrate on the optical intra- and inter-subband transitions mediated by static scatterers such as interface defects, Coulombic donors and alloy disorder, and we show that the dependance of the FCA upon the photon energy is significantly modified by the polarization of the electromagnetic wave. Along the same line, we compare the different FCA magnitudes for two  different polarization configurations in a two-dimensional heterostructure. Notice that in a QCL with in-plane light polarization, the electromagnetic wave propagates through an inhomogeneous medium and thus an absorption coefficient cannot be defined as it is usually done for bulk materials or QCLs with standard light polarization. For that reason we shall evaluate the energy loss rate associated with intra- and inter-subband oblique transitions and not express it in terms of absorption coefficient.

\section{Model}

We consider the active region of a QCL based on an asymmetrical 26/3.1/12.6~nm double quantum well design (DQW) where the wide well is located on the left hand side of the structure. The conduction band offset is set to $V_b=$ 360~meV corresponding to the In$_{0.53}$Ga$_{0.47}$As/GaAs$_{0.51}$Sb$_{0.49}$ ternary system \cite{deutsch2010}. The carrier effective masses are $m^*=$ 0.045$m_0$ in the barrier material and 0.043$m_0$ in the well material respectively. The DQW contains few carriers with an electronic sheet density equal to $n_e=$ 2.17$\times$10$^{10}$~cm$^{-2}$. This structure supports 9 bound states $E_n$ for the $z$ motion where $n$ is the subband index. 
In the absence of disorder and if nonparabolicity is neglected, the two-dimensional eigenstates of the ideal DQW can be written as: 
\begin{align}
 \langle \vec{\rho},z | n,\vec{k} \rangle = \chi_n(z)\frac{1}{\sqrt{S}} e^{i\vec{k}\cdot\vec{\rho}} \\
 \varepsilon_{n\vec{k}}=E_n+\frac{\hbar^2k^2}{2m^*}
 \label{equa1}
\end{align}
where $\vec{\rho}=(x,y)$ is the in-plane position, $\vec{k}$ the two-dimensional wavevector and $S=$ 200$\times$200~nm$^2$ the sample area.

In the presence of static scatterers, the corresponding envelope function Hamiltonian is: 
\begin{equation}
 H=\frac{p^2}{2m^*}+V_b(z)+V_{\mathrm{dis}}(\vec{\rho},z)
 \label{equa2}
\end{equation}
where $V_{\mathrm{dis}}(\vec{\rho},z)$ is the electron potential due to either interface roughness, Coulombic donors or alloy disorder.

\begin{figure}
 \includegraphics{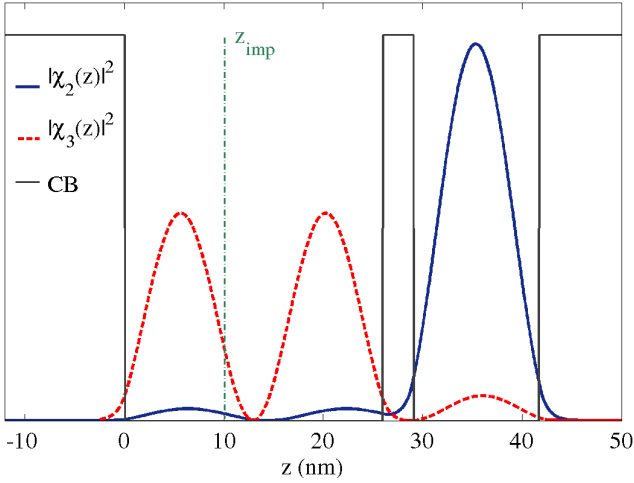}
 \caption{(Color online) Conduction band profile for the 26/3.1/12.6~nm DQW and squared modulus of the $\chi_2$ (blue solid line) and $\chi_3$ (red dashed line) envelope wavefunctions. The vertical green dashed-dotted line represents the donors plane at the position $z_{\mathrm{imp}}=$ 10~nm. The interface defects are located in the two inner interfaces of the structure, i.e, at $z_0=\{$26~nm, 29.1~nm$\}$}.
\label{fig1}
\end{figure}

Fig.~\ref{fig1} shows the conduction band profile of the DQW structure and the squared modulus of the envelope wavefunctions $\chi_2$ and $\chi_3$. 
The interface defects are modeled by Gaussian protrusions of the barrier in the well (repulsive defects) or well in the barrier (attractive defects) \cite{carosella2010}. The Gaussian defects are characterized by their in-plane extension $\sigma$ and are introduced in the two inner interfaces (labelled $z_0$, see Fig.~\ref{fig1}) of the structure with a fractional coverage of the surface $f=\pi\sigma^2n_{def}$ where $n_{def}=n_{att}+n_{rep}$ is the areal concentration of defects.
The Coulombic donors sit on a single doping plane located at $z_{\mathrm{imp}}=$ 10~nm from the left hand side of the wide well  (see Fig.~\ref{fig1}) with an areal concentration equal to $n_e$ corresponding to a distribution of 8 impurities on the $Oxy$ plane.
Using the Virtual Crystal Approximation (VCA) \cite{bastard1996}, we define the alloy disorder potential as a delta-potential characterized by its effective strength $\Delta V$ and the volume of the VCA unit cell $\Omega_0$.

In the following, we study the optical intra- and inter-subband transitions mediated by the static disorder of the structure. If the electric field of the electromagnetic wave is polarized in the layer $Ox$ ($Oy$), the light-matter coupling is described by the dipolar matrix element $\langle f|p_{\bot}|i\rangle$ where $p_{\bot}=p_x$ ($p_y$). Thus, for an electromagnetic wave with an angular frequency $\omega$, the energy loss rate associated with the disorder-assisted transitions $|n\vec{k}\rangle \rightarrow |m\vec{k'}\rangle$ is given by: 
\begin{multline}
 P_{nm}(\omega)=\frac{\pi e^2F_{\mathrm{las}}^2}{m^{*2}\omega} \sum_{\vec{k},\vec{k'}} \left( f_{n\vec{k}} - f_{m\vec{k'}} \right) \left|\langle \Psi_{m\vec{k'}}| p_\bot   | \Psi_{n\vec{k}}  \rangle   \right|^2 \\
 \times \delta\left( \varepsilon_{m\vec{k'}} - \varepsilon_{n\vec{k}} - \hbar\omega  \right)
 \label{equa3}
\end{multline}
where $F_\mathrm{las}$ is the intensity of the laser field, $\Psi_{n\vec{k}}$ and $\Psi_{m\vec{k'}}$ are the perturbed wavefunctions for the intial and final state respectively, and $f_{n\vec{k}}$ and $f_{m\vec{k'}}$ are their respective occupation functions. On account of the low carrier concentration and the relatively large electron temperature ($T=$ 100~K), we assume a thermalized distribution inside the initial $E_n$ subband and take the occupation functions as Boltzmann distributions. 

The perturbed wavefunctions $\Psi_{n\vec{k}}$ and $\Psi_{m\vec{k'}}$ are evaluated by expanding the electronic states to the first order in $V_\mathrm{dis}$ following the perturbative approach of Ref.~\onlinecite{carosella2012}. Then the dipolar matrix element of Eq.~\ref{equa3} reduces to: 
\begin{equation}
 \left|\langle \Psi_{m\vec{k'}}|p_\bot|\Psi_{n\vec{k}}\rangle \right|^2= \frac{\left|\langle m\vec{k'}|V_{\mathrm{dis}}|n\vec{k}\rangle\right|^2 }{\omega^2}
 \left( k_\bot - k_\bot'    \right)^2
 \label{equa4}
\end{equation}
where $k_\bot$=$k_x (ky)$. By averaging over the defects/donors positions, the squared modulus of Eq.~\ref{equa4} becomes proportional to the number of defects/donors and for the interface defects, Coulombic donors and alloy disorder respectively, the analytical development of Eq.~\ref{equa3} gives: 
%
\begin{multline}
   P_{nm}^{\mathrm{def}}(\omega)=\frac{\pi e^2F_{\mathrm{las}}^2\sigma^4V_b^2}{\omega^3\hbar^2m^*}n_e(1-e^{-\beta\hbar\omega})
 \\\times\Lambda_{nm}^{def}
\int_0^{2\pi}d\theta \int_0^\infty du e^{-u}e^{-\sigma^2q_{nm}^2(u)}q^2_{nm}(u)
\label{Pdef}
\end{multline}
\begin{multline}
   P_{nm}^{\mathrm{imp}}(\omega)=\frac{e^6F_{\mathrm{las}}^2 n_{\mathrm{imp}}}{16\omega^3\hbar^2m^*(\varepsilon_0\varepsilon_r)^2}n_e(1-e^{-\beta\hbar\omega}) \\ \times
 \int_0^{2\pi}d\theta \int_0^\infty du e^{-u} \left| \Lambda_{nm}^{\mathrm{imp}}(q_{nm(u)})  \right|^2 \\
 \label{Pimp}
\end{multline}
\begin{multline}
    P_{nm}^{\mathrm{alloy}}(\omega)=\frac{e^2F_{\mathrm{las}}^2 x(1-x)\Delta V^2 \Omega_0}{\omega^3\hbar^2m^*}n_e(1-e^{-\beta\hbar\omega})\\ \times \Lambda_{nm}^{\mathrm{alloy}} 
\int_0^{2\pi}d\theta \int_0^\infty du e^{-u} q_{nm(u)}^2 
 \label{Palloy}
\end{multline}
%
%
%
with:
\begin{align}
  & q_{nm}(u)=\frac{2m^*}{\beta\hbar^2}\left( 2u+\beta(\hbar\omega-E_{nm}) \right. \nonumber
 \\&\times \left.-2\sqrt{u}\sqrt{(u+\beta(\hbar\omega-E_{nm})}\cos\theta   \right) \\
 \label{qnm} 
&\Lambda_{nm}^{\mathrm{def}}=\sum_{z_0} \left( n_{\mathrm{att}}\left|\int_{z_0-h}^{z_0} dz \chi_n(z)\chi_m(z)\right|^2 \right. \nonumber\\ &+ 
 \left. n_{\mathrm{rep}}\left|\int_{z_0}^{z_0+h} dz \chi_n(z)\chi_m(z)\right|^2 \right) \\
 \label{lambda_def}
%
 &\Lambda_{nm}^{\mathrm{imp}}(q_{nm(u)})=\int dz \chi_n(z)\chi_m(z)e ^{-q_{nm(u)}|z-z_{\mathrm{imp}}|} \\
&\Lambda_{nm}^{\mathrm{alloy}}=\int_{\mathrm{alloy}} dz \chi_n(z)^2\chi_m(z)^2
 \label{lambda_alloy}
\end{align}
and where $u=\frac{\beta\hbar^2k^2}{2m^*}$, $\beta=(k_BT)^{-1}$ and $E_{nm}=E_m-E_n$ is the bare transition energy.

\section{Results and discussion}

We have numerically evaluated the contributions to the energy loss rate by computing Eqs.~\ref{Pdef}-\ref{Palloy} for the DQW structure described above and for $n=2$ and $m=2$ (intra-subband transition) or $m=3$ (inter-subband transition). The fractional coverage of interface defects was set to $f=0.3$ and the defect size to $\sigma=5.6$~nm. Since the electronic wavefunctions are mostly localized in the wells of the structure (see Fig.~\ref{fig1}), we considered the alloy disorder in the  GaInAs wells only, with an alloy fraction of $x=0.53$ and an effective strength of $\Delta V=0.6$~eV. $F_{\mathrm{las}}=1$~kV.cm$^{-1}$

\begin{figure}
 \includegraphics{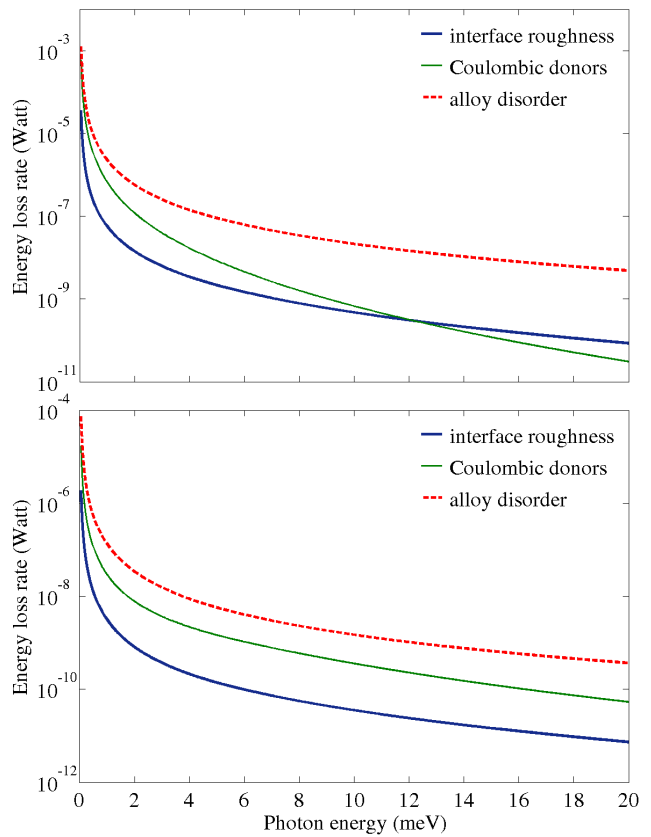}
 \caption{(Color online) Energy loss rate $P_{nm}$ versus the photon energy $\hbar\omega$ for the intra-subband $E_2$ (upper panel) and inter-subband $E_2-E_3$ (lower panel) transitions. $P_{nm}$ is calculated by taking into account either the interface roughness (blue solid line), donors (green solid line) or alloy disorder (red dashed line). The electromagnetic wave is polarized in the layer plane. $T=$ 100~K.
}
\label{fig2}
\end{figure}

We show in Fig.~\ref{fig2} the energy loss rate versus the photon energy $\hbar\omega$. We compare the situation where either the interface defects, the Coulombic donors or the alloy disorder are the scatterers and where the electromagnetic wave is polarized along an in-plane direction. As expected from Eqs.~\ref{Pdef}-\ref{Palloy}, the plots of Fig.~\ref{fig2} display a dependance of the energy loss rate upon the photon energy that goes like $\omega^{-p}$ where $p>0$ and depends on the type of scatterer. In this configuration, the conduction states are extended in the $x$ and $y$ directions, the carrier free motion then occurs in the same plane as the electric field direction. Thus, the absorption processes can be reliably estimated by the semi-classical description assuming that the carriers are accelerated by the electric force $-e\vec{F}_{\mathrm{las}}$. As a consequence, the $\omega^{-p}$ ($p\sim2-3$) bulk behavior characteristic of the Drude-like approach \cite{ashcroft1976,walukiewicz1979}, is recovered in this two-dimensional system for this in-plane polarization configuration. 

We also note that the intensity of the intra-subband energy loss rate is of one order of magnitude larger than the inter-subband corresponding one. This can be readily explained by the larger wavefunction overlap in the factor $\Lambda_{nm}$ of Eqs.~\ref{Pdef}-\ref{Palloy} for the intra-subband events, or, stated differently, by the fact that the inter-subband dipolar matrix element $\langle n\vec{k} | p_\bot | m\vec{k'} \rangle=\delta_{n,m}\delta_{\vec{k},\vec{k'}}\hbar{k_\bot}$ leads to inter-subband oblique transitions that are ``doubly forbidden`` in an ideal system, while the intra-subband transitions are forbidden only by wavevector conservation.
\begin{figure}
 \includegraphics{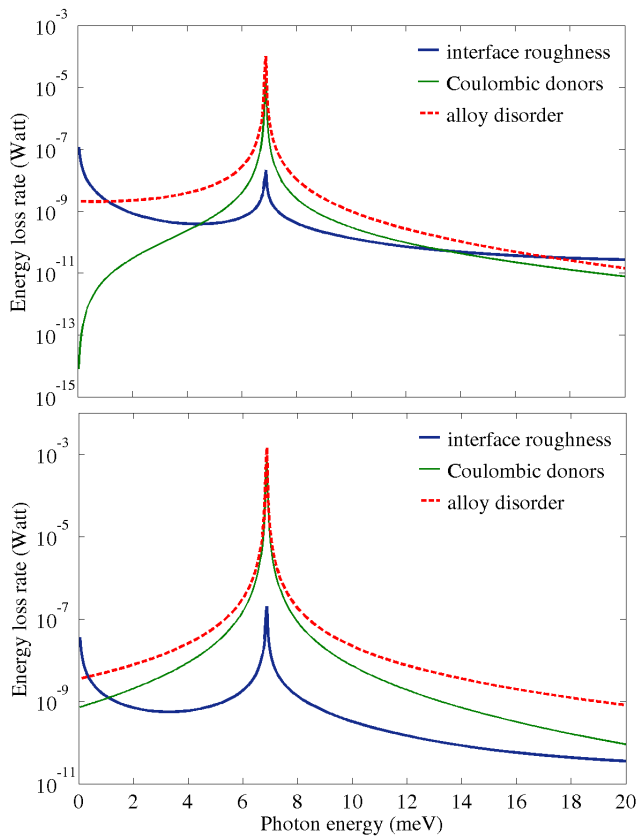}
 \caption{(Color online) Energy loss rate $P_{nm}$ versus the photon energy $\hbar\omega$ for the intra-subband $E_2$ (upper panel) and inter-subband $E_2-E_3$ (lower panel) transitions. $P_{nm}$ is calculated by taking into account either the interface roughness (blue solid line), donors (green solid line) or alloy disorder (red dashed line). The electromagnetic wave is polarized along the growth axis. $E_3-E_2=$ 6.8~meV. $T=$ 100~K.
}
\label{fig3}
\end{figure}
For comparison, we computed the energy loss rate in the $z$-polarization configuration, following the quantum mechanical treatment of FCA established in Ref.~\onlinecite{carosella2012}. The results are shown in Fig.~\ref{fig3} for the two types of transition and for the same material and disorder parameters used in Fig.~\ref{fig2}. Instead of a divergence at $\omega\rightarrow0$, the plots of Fig.~\ref{fig3} display a strong increase when $\hbar\omega$ approaches the bare transition energy $E_3-E_2$. This feature is due to the virtual intermediate coupling in subband $E_3$ to get a nonzero absorption for these oblique transitions, as explained in Ref.~\onlinecite{carosella2012}.
By comparing the intra- and inter-subband contributions, we get the reverse trend than in the in-plane case: for a given scattering source and in the $z$-polarization case, the intra-subband process displays a weaker energy loss rate than the inter-subband one. This contrast is due to the ``doubly forbidden`` nature of the intra-subband oblique transition \cite{carosella2012}, i.e., contrary to the in-plane configuration, the associated dipolar matrix element $\langle n\vec{k}|p_z|m\vec{k'}\rangle=\langle n |p_z|m\rangle\delta_{\vec{k},\vec{k'}}$ leads to intra-subband transitions that are doubly forbidden in an ideal system and inter-subband oblique transitions that are forbidden only by wavevector conservation.
This feature also explains that, far from the resonance, the intra-subband energy loss rate for the $z$-polarization is at least two orders of magnitude weaker than the corresponding one in the in-plane configuration, while the inter-subband events lead to similar energy loss rate magnitude for both polarizations.  

Finally, Figs.~\ref{fig2} and \ref{fig3} show that FCA is a more efficient loss mechanism when the elastic scattering is dominated by the alloy disorder contribution in ternary two-dimensional heterostructures. Notice that the alloy disorder has already been demonstrated \cite{bastard1983} as being an efficient scattering mechanism in this kind of material characterized by an indium concentration ($x$=0.53) that is close from the one ($x$=0.5) that maximizes the alloy scattering efficiency. 

\section{Conclusion}

We have evaluated FCA in cascade structures associated with static scatterers when the wave propagates along the growth axis. For such a polarization no absorption coefficient can be defined. The computation of the energy loss rate allows, for a given intensity of the electromagnetic field, to compare the absorption for the in-plane polarization to the one for the polarization along the growth axis. We found that the electromagnetic waves that propagate along the growth axis are more absorbed than those that propagate in the layer plane and that their absorption can be described in terms of a Drude-like decrease: $P(\omega)\approx\omega^{-p}$, where $p$ is positive and depends on the nature of the scattering mechanism. This result strikingly contrasts with the case of QCL's losses due to FCA. Moreover, when the electromagnetic wave is in-plane polarized, the intra-subband absorption is found significantly larger than that associated with oblique inter-subband transitions, because these ones are more forbidden than the inter-subband transitions. This result is the opposite of the one evaluated for the standard QCL polarization.
\\

\begin{acknowledgements}
G.B. thanks the Technical University Vienna for hospitality. Discussions with A. Wacker, G. Strasser and K. Unterrainer are gratefully acknowledged. 
\end{acknowledgements}

\end{document}